\documentclass{appolb}
\usepackage{epsfig}
\usepackage{eqnarray,amsmath}
\usepackage{color}
\usepackage{caption}
\usepackage{chemscheme}
\begin{document}
% \eqsec  % uncomment this line to get equations numbered by (sec.num)
\title{Bayesian analysis for a new class of hybrid EoS models using mass and radius data of compact stars
\thanks{Presented at the International Conference on Critical Point and Onset of Deconfinement (CPOD'2016), May, 29 -- June, 5, 2016, University of Wroclaw}%
% you can use '\\' to break lines
}

\author{A.~Ayriyan$^1$, D.~E.~Alvarez-Castillo$^2$, S.~Benic$^{3,4}$, D.~Blaschke$^{2,5,6}$, H.~Grigorian$^{1,7}$, S.~Typel$^8$
\address{
%\begin{center}
$^1$~Laboratory of Information Technologies, JINR, RU-141980 Dubna\\
$^2$~Bogoliubov Laboratory for Theoretical Physics, JINR, RU-141980 Dubna\\
$^3$~Department of Physics, The University of Tokyo, JP-113-0033, Tokyo\\
$^4$~Physics Department, The University of Zagreb, HR-10000, Zagreb\\
$^5$~Institute of Theoretical Physics, Wroclaw University, PL-50-204 Wroclaw\\
$^6$~National Research Nuclear University (MEPhI), RU-115409 Moscow\\
$^7$~Department of Theoretical Physics, Yerevan State University, AM-0025 Yerevan\\
%$^7$~Theoretical Physics Department, Yerevan State University, AM-0025 Yerevan\\
$^8$~GSI Helmholtzzentrum f\"ur Schwerionenforschung GmbH, DE-64291 Darmstadt
%\end{center}
}
}

%\author{Put here the name(s) of the Author(s)
%\address{and affiliation}
%\and
%the Name(s) of other Author(s)
%\address{and their affiliation}
%}

\maketitle
\vspace{-1cm}
\begin{abstract}
We present a Bayesian analysis for a new class of realistic models of two-phase equations of state (EoS) for hybrid stars and demonstrate that the observation of a pair of high-mass twin stars would have a sufficient discriminating power to favor hybrid EoS with a strong first order phase transition over alternative EoS.
Such a measurement would provide evidence for the existence of a critical endpoint in the QCD phase diagram.
\end{abstract}

\section{Introduction}
In this contribution we present results of 
a Bayesian analysis (BA) performed with a new class realistic models of two-phase equations of state (EoS) for hybrid stars that allows for a broad variety of mass-radius (M-R) sequences upon variation of two EoS parameters \cite{Alvarez-Castillo:2016oln}. 
These include ordinary neutron stars, stable hybrid star branches connected to neutron star ones and branches disconnected from the neutron star ones.
A disconnected hybrid star branch, also called "third family", covers also a mass range of high-mass twin stars \cite{Benic:2014jia,Alvarez-Castillo:2016wqj} characterized by the same gravitational mass but different radii.
For the classification of M-R sequences with hybrid stars, see \cite{Alford:2013aca}.

The~new class of two-phase EoS is characterized by three main features:
\vspace{-3mm}
\begin{itemize}
  \item[(1)] stiffening of the nuclear EoS at supersaturation densities due to quark exchange effects (Pauli blocking) between hadrons, modelled by an excluded volume correction, \vspace{-3mm}
  \item[(2)] stiffening of the quark matter EoS at high densities due to multiquark interactions and \vspace{-3mm}
  \item[(3)] possibility for a strong first order phase transition with an early onset and large density jump.
\end{itemize}
\vspace{-3mm}

The third feature results from a Maxwell construction for the possible transition from the nuclear to a quark matter phase and its properties depend on the two parameters determining the properties (1) and (2), respectively. Varying these two parameters one obtains a class of hybrid EoS that yields solutions of the Tolman-Oppenheimer-Volkoff (TOV) equations for sequences of hadronic and hybrid stars in the mass-radius diagram which cover the full range of patterns according to the Alford-Han-Prakash classification 
\cite{Alford:2013aca}.

We will use the BA to demonstrate that the observation of a pair of high-mass twin stars would have a sufficient discriminating power to favor hybrid EoS with a strong first order phase transition over alternative EoS~\cite{Alvarez-Castillo:2016oln}.

\section{New class of quark-hadron EoS for hybrid stars}

In this study hybrid neutron stars that are composed of hadronic matter and might undergo a phase transition to quark matter in their cores if parameter values of the models physically allow for~it. 

For the hadronic part of the neutron star EoS we consider here the density dependent relativistic meanfield  EoS named "DD2F{-}" which is slightly softer than "DD2" \cite{Typel:2009sy} as it fulfils the flow constraint from heavy-ion collision experiments \cite{Danielewicz:2002pu} and has a soft symmetry energy. 
The excluded volume correction is applied to hadronic matter at suprasaturation densities and has the effect of stiffening the EoS without modifying any of the experimentally well constrained properties below and around the saturation density $n_{\rm sat}=0.16$ fm$^{-3}$.
%, the density in the interior of atomic nuclei.
 We introduce the  
the available volume fraction $\Phi_N$ for the motion of nucleons at a given density $n$ as~\cite{Typel:2016srf}
$$
    \Phi_N=\left\{
                \begin{array}{lll}
                  1~,& \textmd{if} &n \leq n_{\rm sat}\\
                  \exp[-{v\vert v \vert}(n-n_{\rm sat})^{2}/2]~, & \textmd{if} &n > n_{\rm sat}~,
                \end{array}
              \right.
              \label{vex}
$$
with $v=16\pi r_N^3/3$ as the van-der-Waals excluded volume corresponding to a nucleon hard-core radius $r_N$.
The available volume fraction $\Phi_N$ is introduced for the motion of nucleons at a given density $n$ as the van-der-Waals excluded volume corresponding to a nucleon hard-core radius $r_N$. In this work the dimensionless parameter $p=10\times v[{\rm fm}^3]$ is introduced, taking values $p=0,5,10, \dots, 80$. 

The quark matter EoS models in the high density phase is obtained from a NJL model with multiquark interactions \cite{Benic:2014jia,Benic:2014iaa} where the Lagrangian for two quark flavors, $q=(u,d)$ is defined within the mean-field approximation the thermodynamic potential is 
%\vspace{-2mm}
\begin{eqnarray}
\Omega_{{\color{white}f}} &=& U+ \sum_{f=u,d}\Omega_f(M_f,T,\tilde{\mu}_f)-\Omega_0~,
\nonumber \\
%\end{eqnarray}
%\vspace{-1cm}
%\begin{eqnarray}
\Omega_f &=&-2 N_c\int \frac{d^3 p}{(2\pi)^3}
\Big\{E_f + T\ln[1+e^{-\beta(E_f-\tilde{\mu}_f)}] + T\ln[1+e^{-\beta(E_f+\tilde{\mu}_f)}]\Big\},\nonumber\\
U &=& \frac{2g_{20}}{\Lambda^2}\left[(\phi_u^2+\phi_d^2) - \eta_2(\omega_u^2 + \omega_d^2)\right]
+ \frac{12g_{40}}{\Lambda^8}\left[(\phi_u^2+\phi_d^2)^2 - \eta_4(\omega_u^2+\omega_d^2)^2\right].
\nonumber%\\
\end{eqnarray}
We neglected the mixing term $g_{22}=0$ \cite{Benic:2014iaa} and  set $\eta_2=0.08$.
The parameter $\eta_4$ is the dimensionless scaled coupling strength for the 8-quark interaction in the vector meson channel which determines the stiffness of the quark matter EoS at high densities.
It will be varied in this study as $\eta_4=0,1,2, \dots, 30$.

\begin{figure}[!ht]
\begin{tabular}{ccc}
\includegraphics[width=0.3\linewidth]{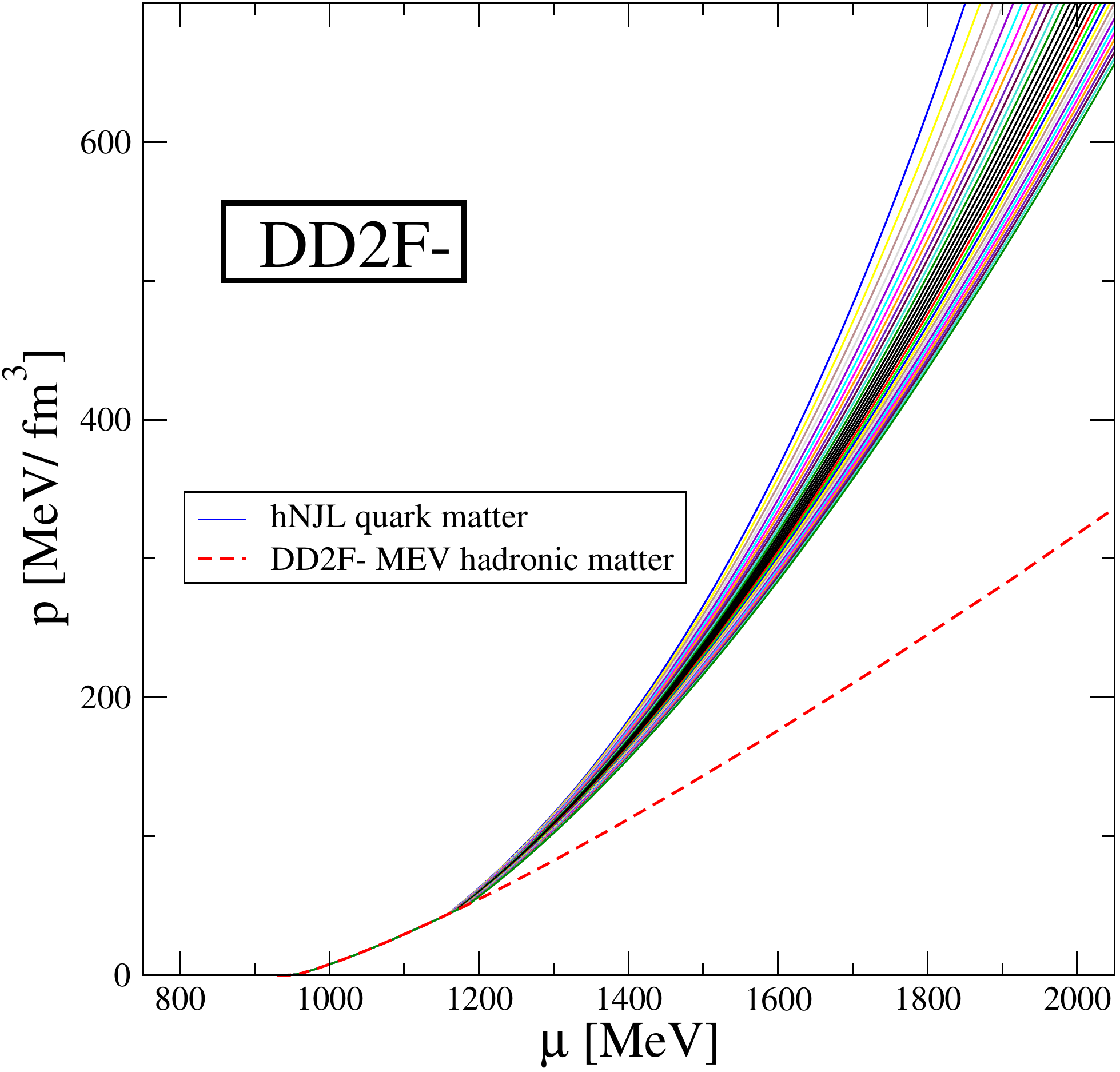} & \includegraphics[width=0.3\linewidth]{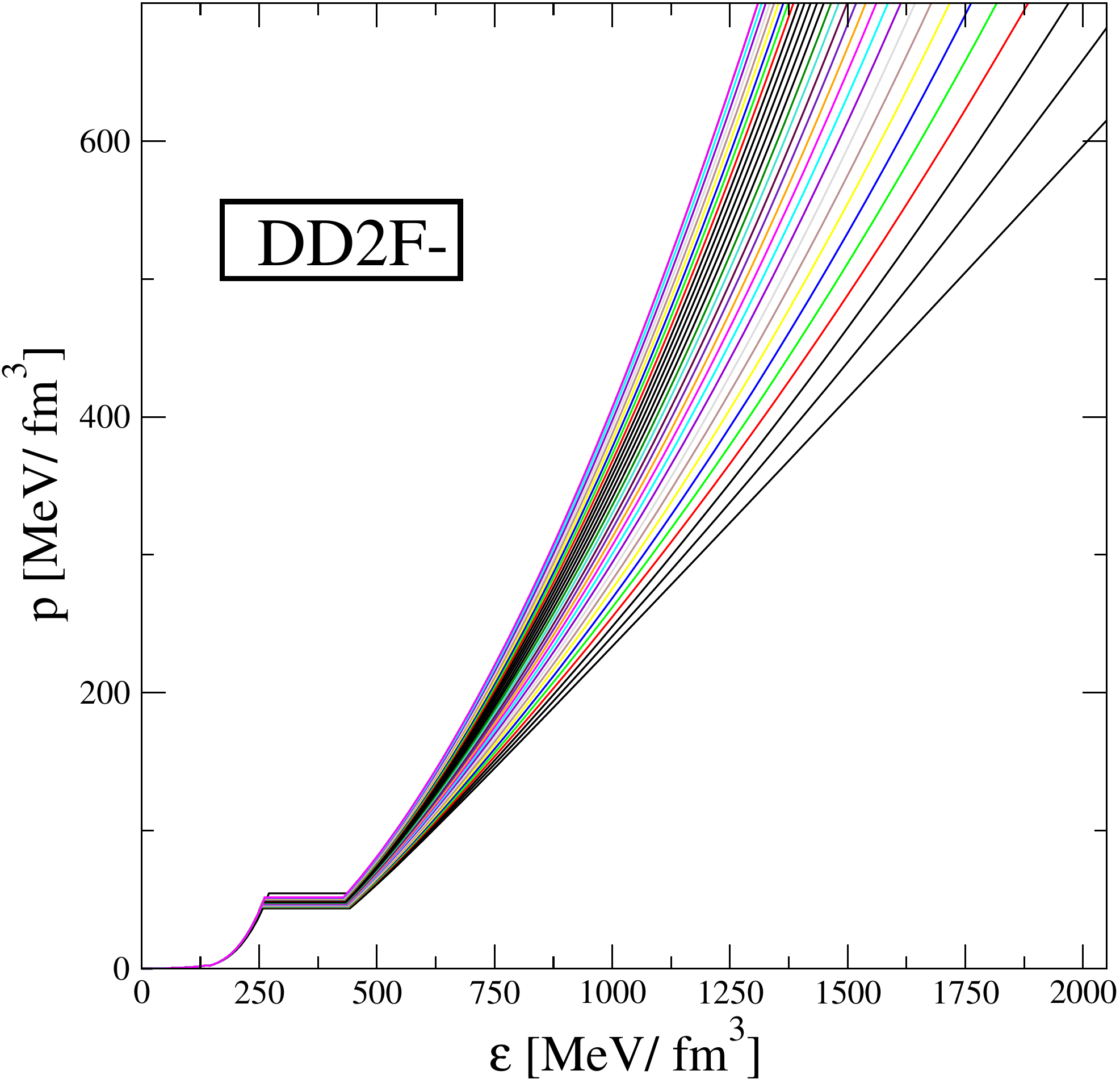} & \includegraphics[width=0.3\linewidth]{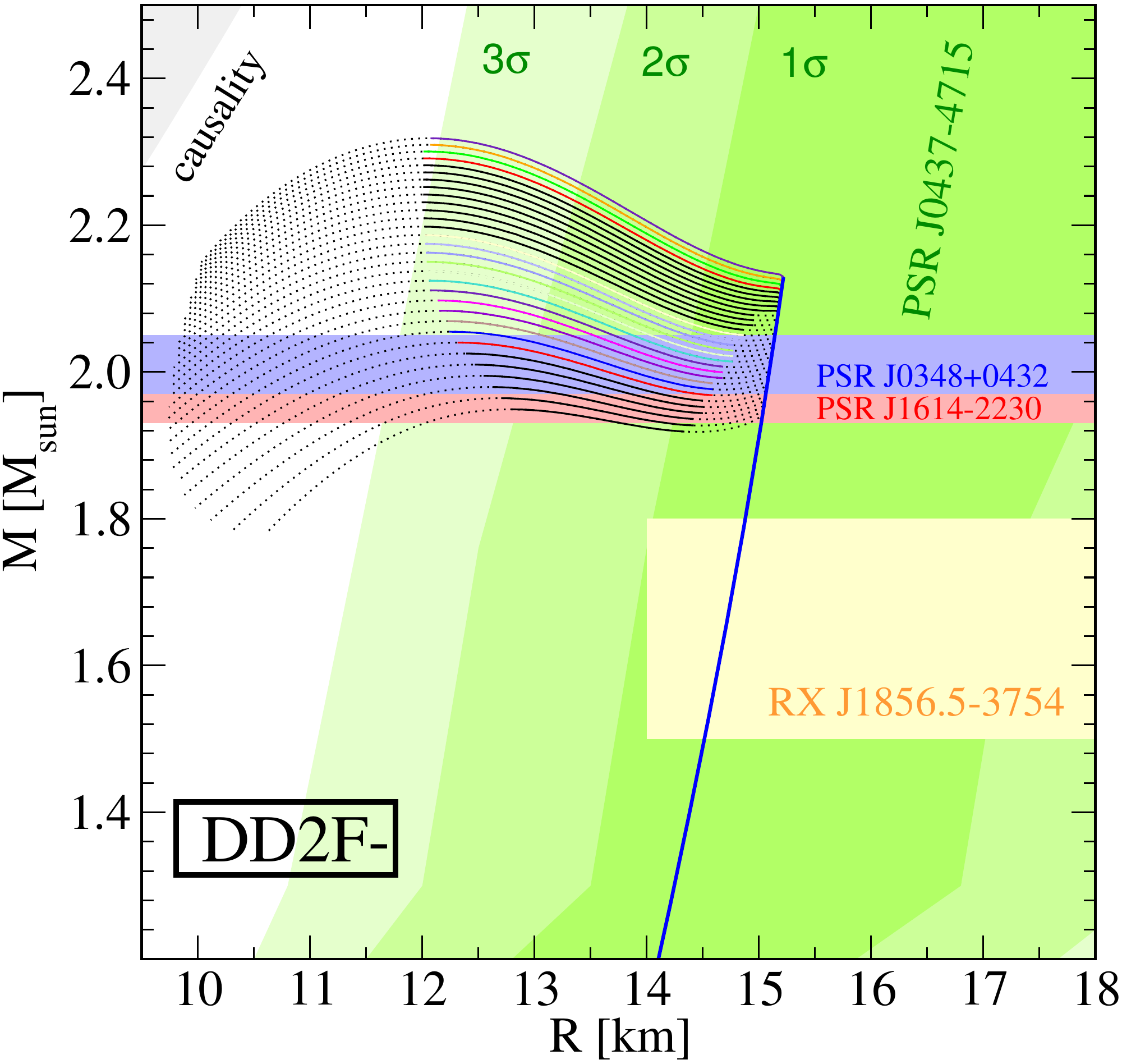}\\
\includegraphics[width=0.3\linewidth]{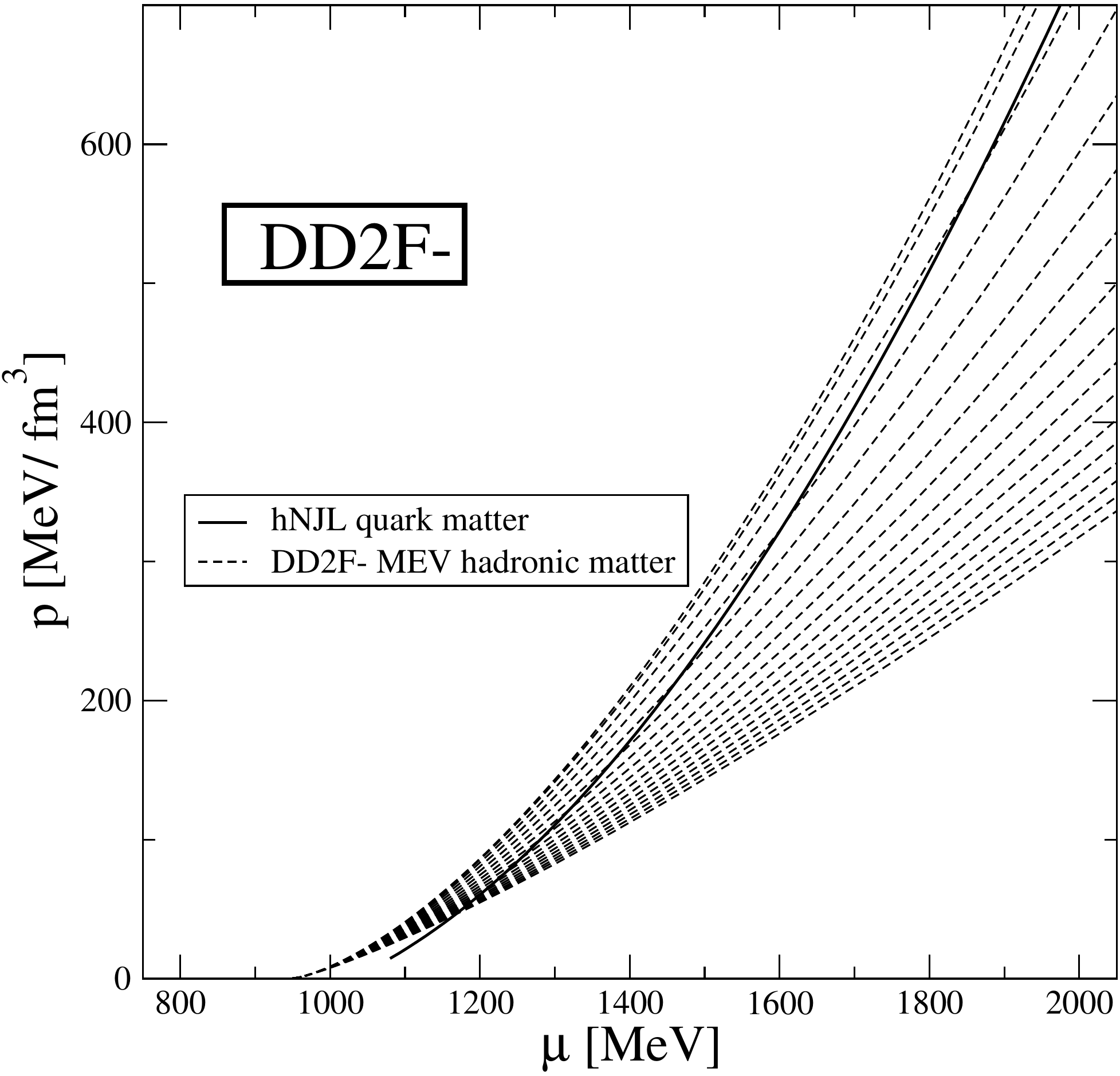} & \includegraphics[width=0.3\linewidth]{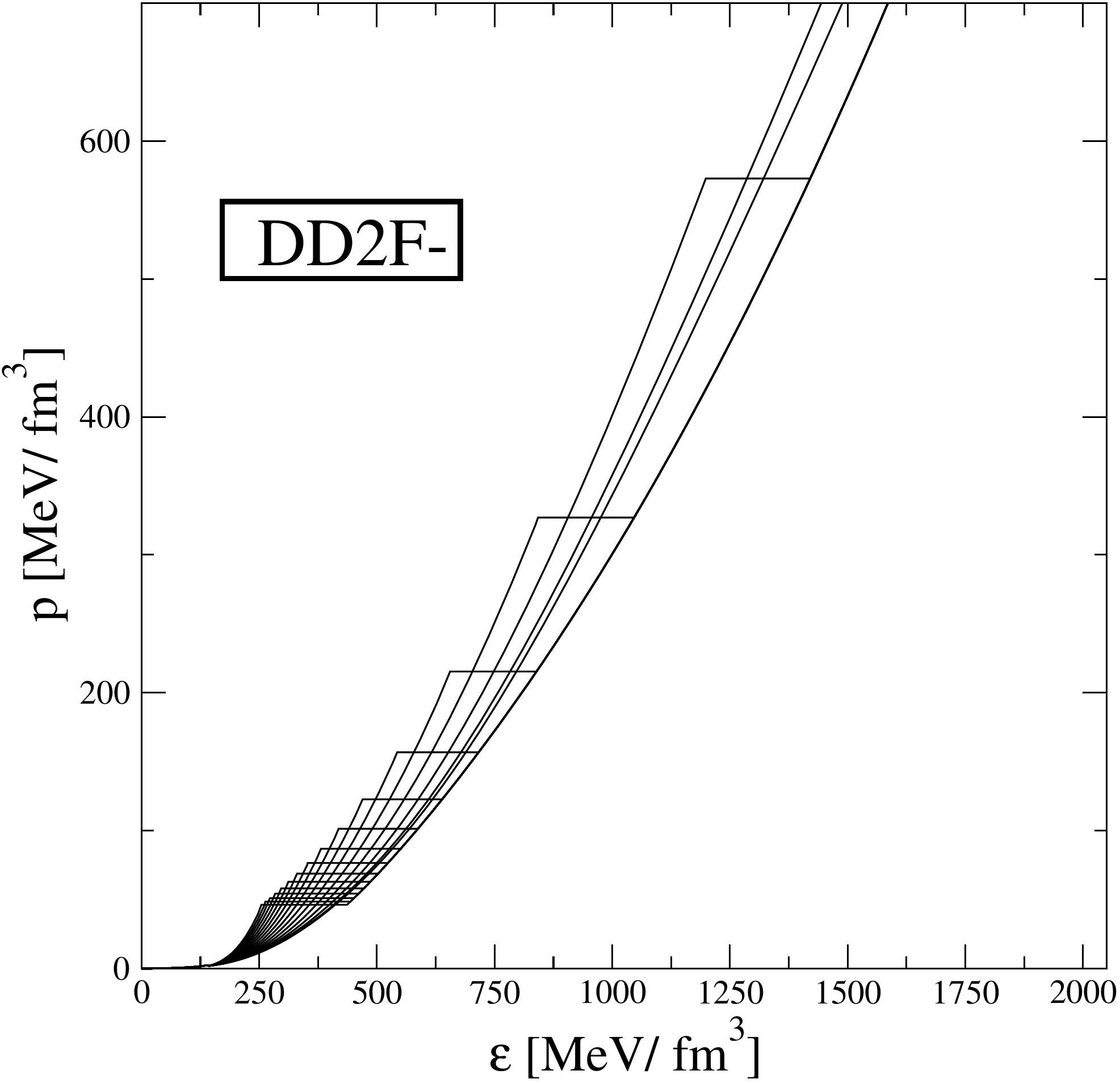} & \includegraphics[width=0.3\linewidth]{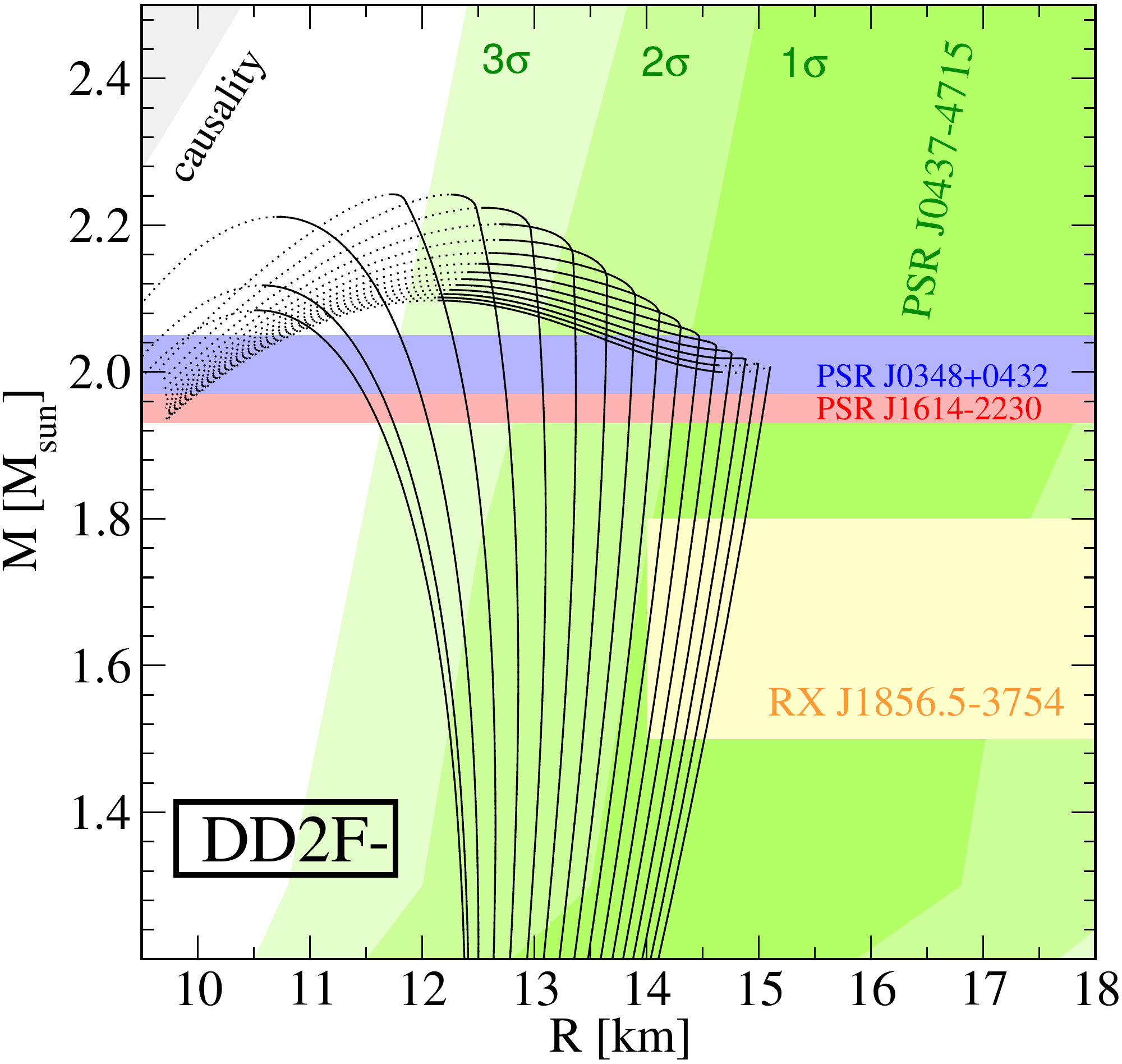}
\end{tabular}
\caption{Variations of the hybrid EoS for the DD2F$^-$ model for fixed hadronic EoS (upper panels)
and for fixed quark EoS (lower panels).
%{Upper row:} The hadronic EoS is kept fixed while the quark EoS is allowed to vary for the parameters $\eta_4=0,1,2, \dots, 30$. 
%{Lower row:} The quark EoS is fixed whereas the hadronic EoS takes the values $p=0,5,10, \dots, 80$. For all these models 
The EoS is shown in the left and middle panels, the M-R diagrams in the right ones.
\label{Case_A_n_B} }
\end{figure}

The phase transition in the hybrid EoS models is obtained by a  Maxwell construction and therefore of
first order.
% (the transition point is determined by enforcing the \textit{Gibbs conditions}). 
%Whether the resulting compact star will be a hybrid or a pure hadronic will depend on having these conditions fulfilled at densities lower than the central density of the most massive star of each EoS sequence.
Having defined the hybrid EoS by a pair of parameters $(p,\eta_4)$, the Tolman-Oppenheimer-Volkoff (TOV) equations can be solved in a standard way resulting in a sequence of stars in the M-R diagram for each parameter set.
A systematic analysis is performed by varying the EoS parameters, see  Fig.~\ref{Case_A_n_B}. 
%This has important consequences for the compactness of the neutron star as well as for the possible existence of a disconnected branch of hybrid stars ("third family") in the mass-radius diagram. 

\section{Bayesian analysis for Compact Stars}

BA is considered a powerful technique for model and parameter descrimination. 
The different observables taken into account for the present BA are 
the highest precisely measured masses $M_{{A}}=2.01~M_{\odot}$  for PSR~J$0348+0432$~\cite{Antoniadis:2013pzd} and $M_{{D}}=1.94~M_{\odot}$ for PSR~J$1614-2230$~\cite{Demorest:2010bx,Fonseca:2016tux} with an error band $\Delta M_{{A}}=\Delta M_{{D}}=\pm0.04~M_{\odot}$
as well as the radius determination $R_{B}=15.5$ km with $\sigma_{R_{B}} = 1.5$ km  by Bogdanov \cite{Bogdanov:2012md} for the nearest millisecond pulsar PSR J0437-4715. 
These constraints are shown by the colored bands in the right panels of Fig.~\ref{Case_A_n_B}.
They were already included in our earlier BA works \cite{Blaschke:2014via, Alvarez-Castillo:2014nua, Alvarez-Castillo:2015via, Ayriyan:2015kit, Alvarez-Castillo:2016oln} where more details can be found. 

In order to provide guidance for strategies of future observational programmes we employ fictitious radius measurements. 
We assume that the radii $R_{A}$ and $R_{D}$ of both above mentioned high-mass pulsars could be measured with a resolution characterized by the statistical uncertainty $\sigma_{R_{A}}$ and $\sigma_{R_{D}}$, respectively.

%\begin{center}
\begin{figure}[!ht]
\begin{center}
\includegraphics[width=0.5\linewidth]{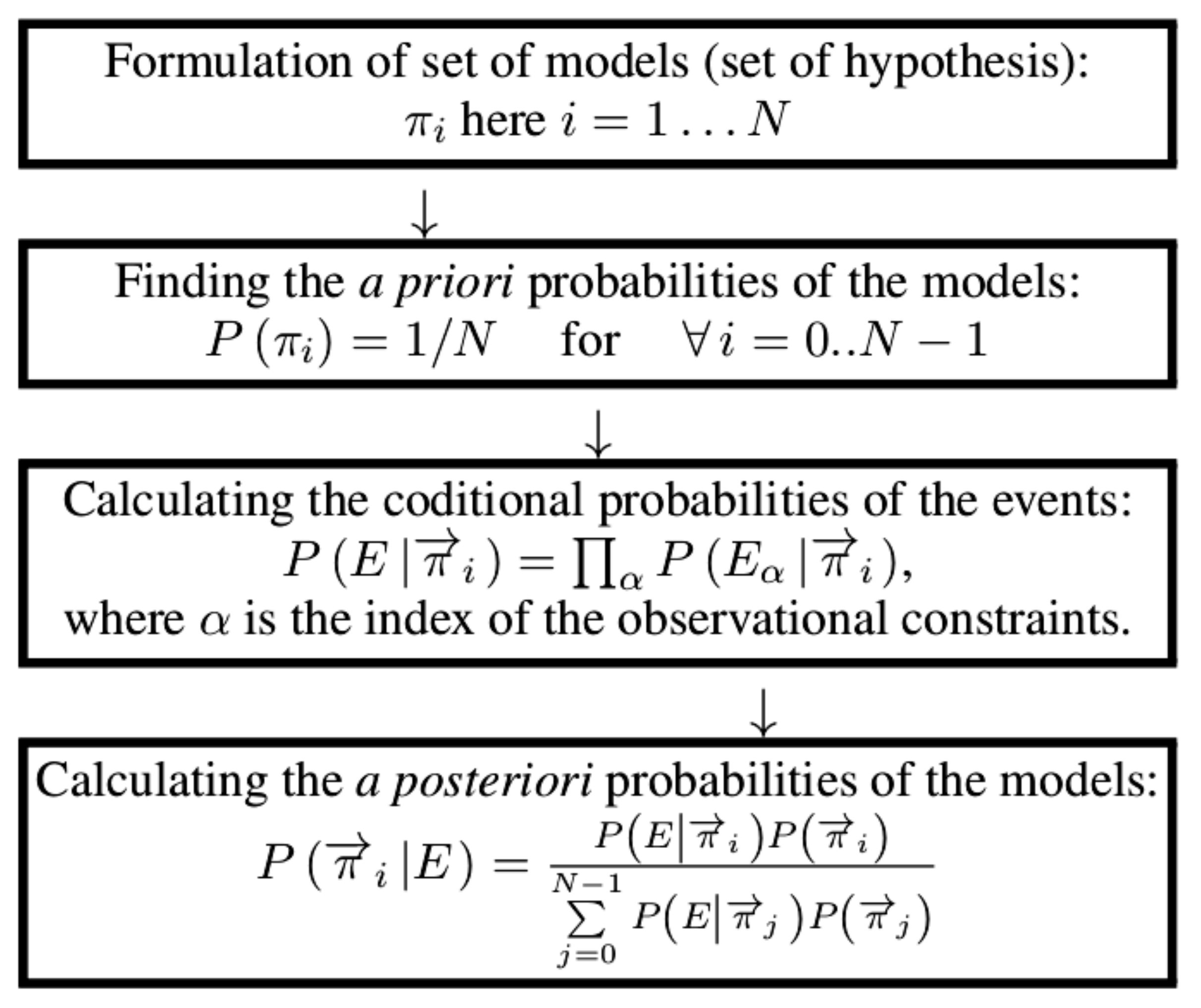}
\end{center}
\captionof{scheme}{\label{ba:scheme}
The scheme of the Bayesian analysis technique for the EoS model using astrophysical data.}
\end{figure}
The vector of free parameters which correspond to all the possible models with or without nuclear to quark matter phase transition is defined as $\overrightarrow{\pi}_{i}=\{ p_{(k)},\eta_{4(l)}\}$, where $i=0\dots N-1$ with $N=N_{1}\times N_{2}$ such that $i=N_{2}\times k+l$
and $k=0\dots N_{1}-1$, $l=0\dots N_{2}-1$, with $N_{1}$ and $N_{2}$ being the total number of parameters $p_{(k)}$ and $\eta_{4(l)}$, respectively.
After integration of the TOV equations each EoS model results in a curve in the M-R diagram for which the probability can be determined that this EoS fulfils the chosen observational constraints. 
To this end the conditional probabilities of these constraints are calculated which quantify how probable the observational data are for the assumed EoS model. 
The goal of the BA is to find the set of most probable $\overrightarrow{\pi}_{i}$ matching the above constraints using the BA technique (see scheme~\ref{ba:scheme}). 

\begin{figure}[!ht]
\begin{tabular}{ ccc }
\includegraphics[width=0.3\linewidth]{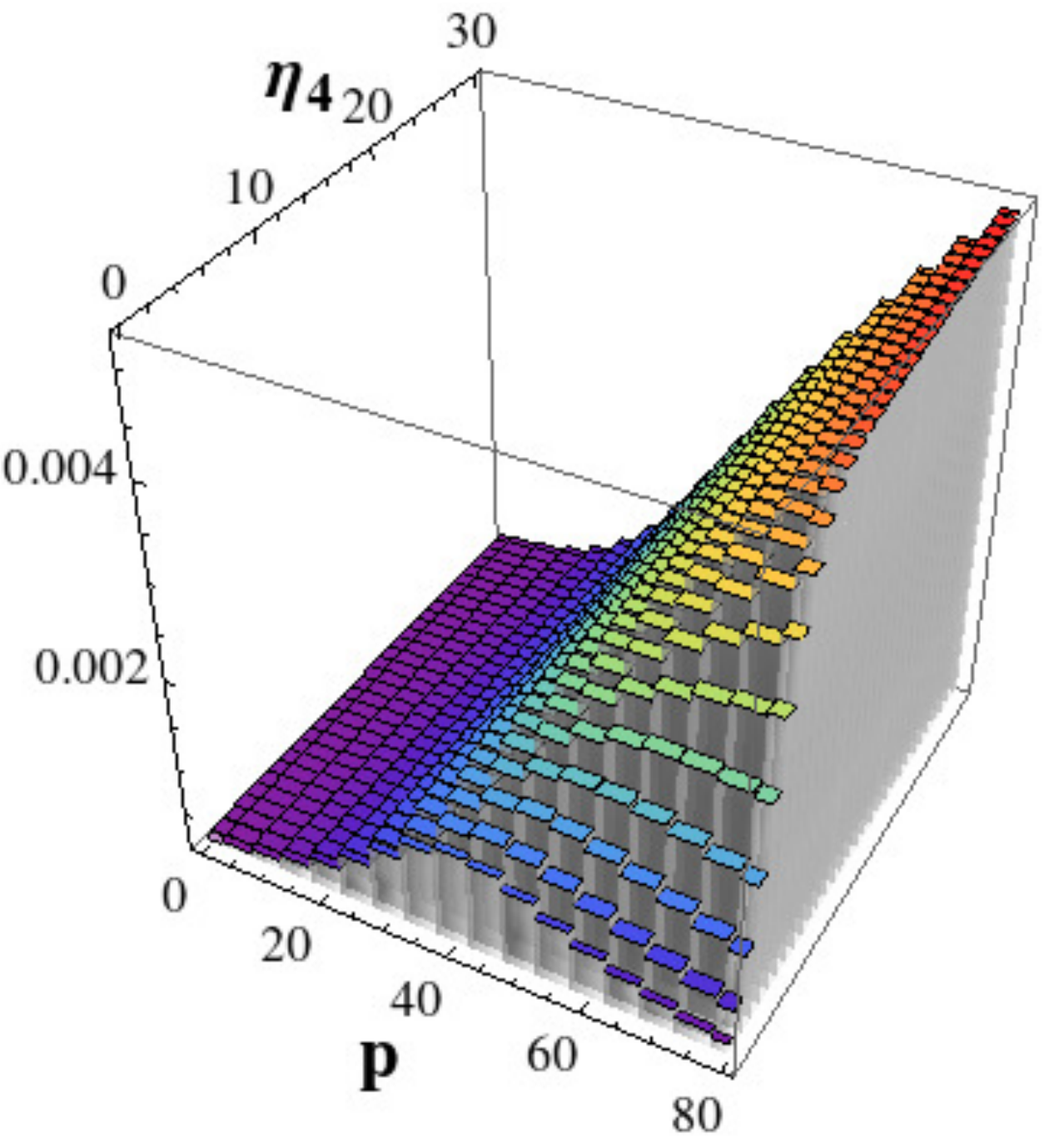} & 
\includegraphics[width=0.3\linewidth]{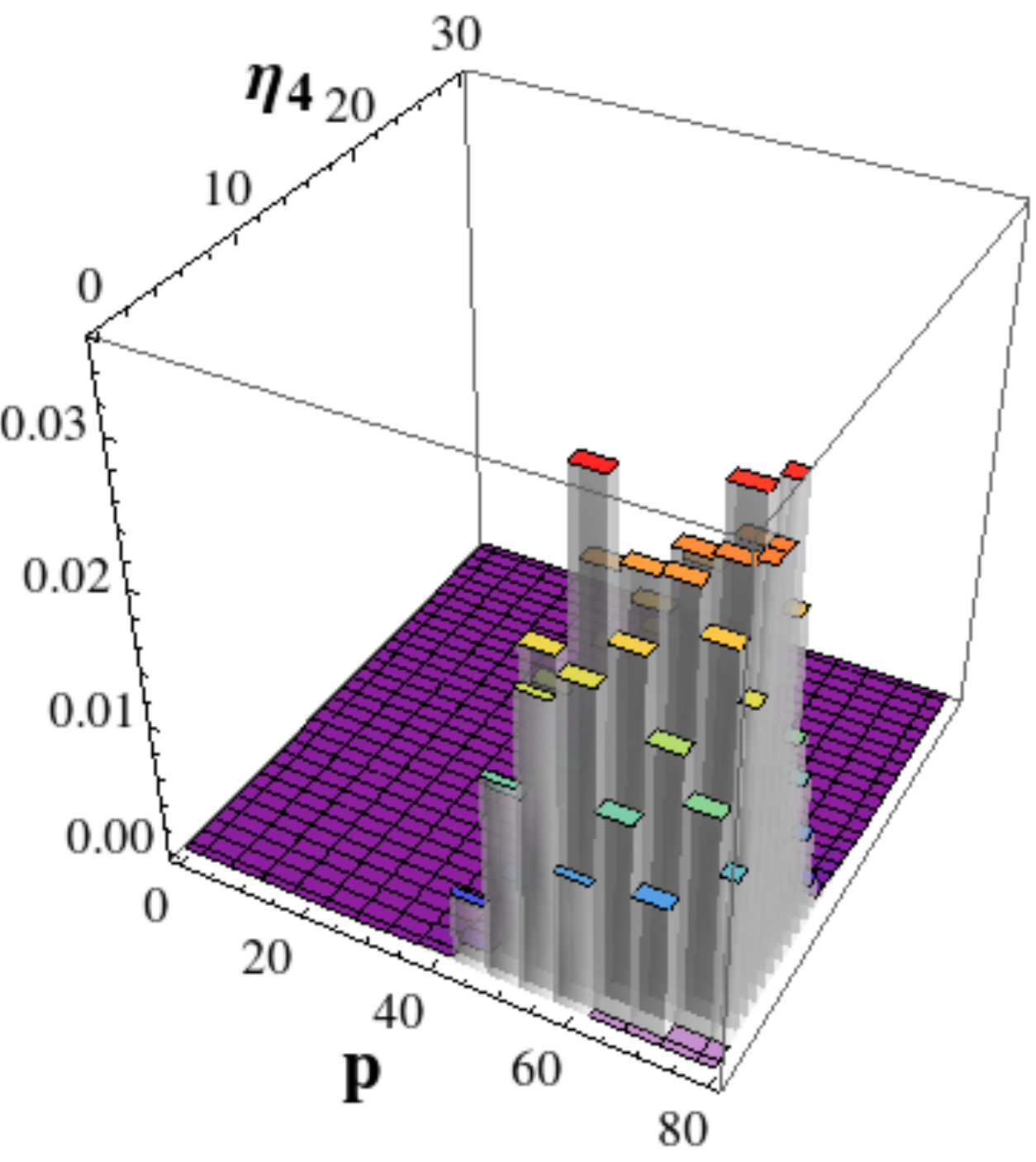}&
\includegraphics[width=0.3\linewidth]{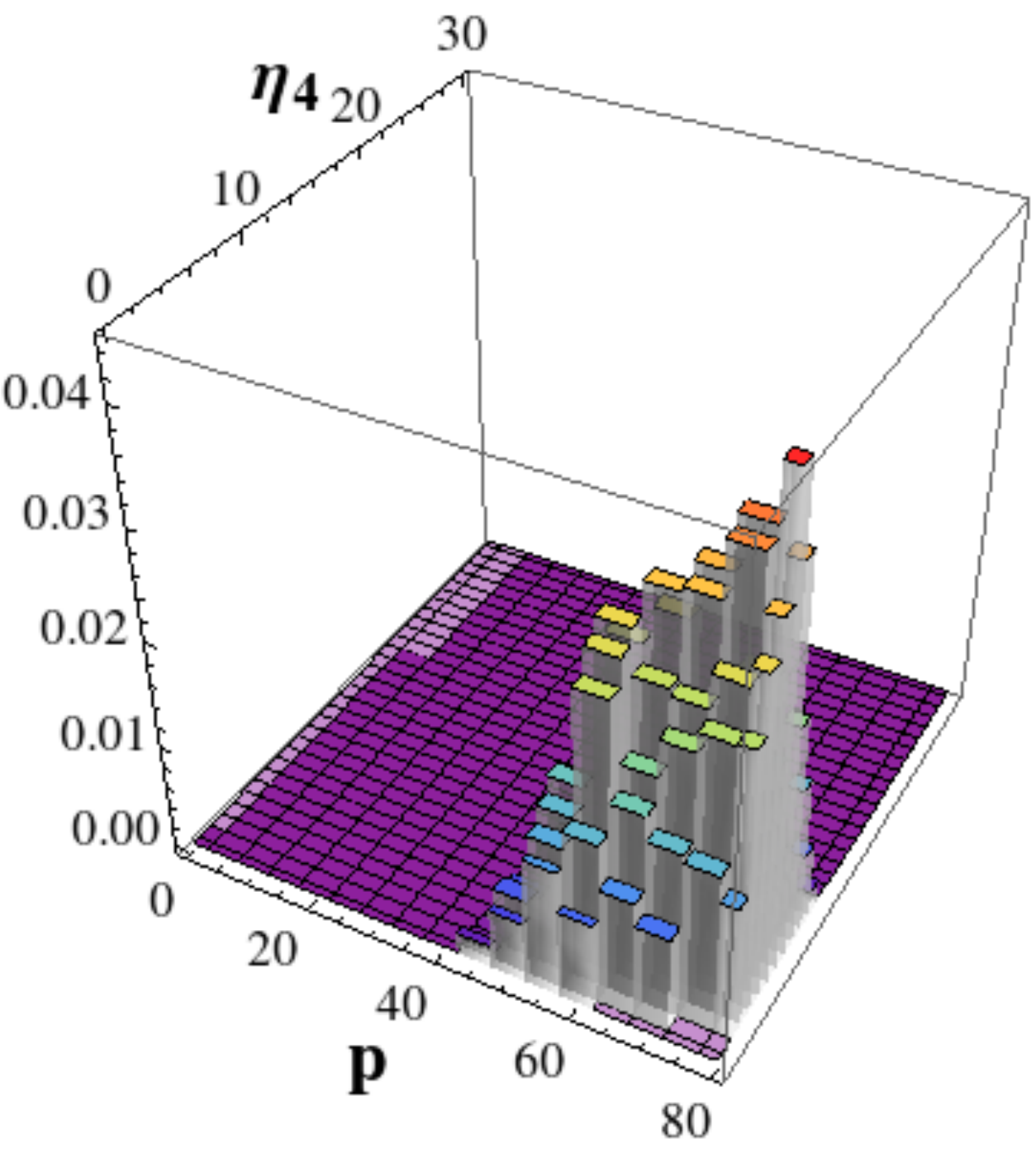}
\end{tabular}
\caption{BA of the probabilities in the EoS model parameter space $\eta_4 - p$ for separate 
mass and radius observation $(M_A,R_B)$ (left panel) compared to fictitious radius measurements 
for known high-mass pulsars as high-mass twins with $(M_A,13~{\rm km})$ and $(M_D,15~{\rm km})$
for $\sigma_R=1.5$ km (middle panel) and $\sigma_R=0.5$ km (right panel).
\label{apost_prob} }
\end{figure}

\section{Results and Conclusions}

In the left panel of Fig.~\ref{apost_prob} we show the BA results when using the mass constraint 
$M_{{A}}$ together with the radius constraint $R_B$.
One can see that existing mass-radius constraints have high selective power for hadronic part of considered EoS models, whereas  for quark phase they have practically no influence to the multiquark interaction parameter. The most probable values of the excluded volume parameter is located in  the ranges $40<p < 80$ at the two-dimensional parameter space (see~fig.~\ref{apost_prob}).

Next we have performed a BA assuming a set of fictitious radii $(R_A,R_D)=(13~{\rm km}, 15~{\rm km})$ for the known masses of the two high-mass pulsars. 
In the middle and right panels of  Fig.~\ref{apost_prob} we present the results for the radius uncertainties 
of 1.5 km and 500 m, respectively.
The results demonstrate that the simultaneous measurement of radii and masses of a pair of high-mass twin stars (here by assuming the possible outcome of radius measurements) could be strongly selective and could have sufficient discriminating power to favor hybrid EoS with a strong first order phase transition over alternative EoS.

The next two steps in the development of the approach are devoted to an improvement of the variability of the dense matter EoS within a two-dimensional parameter space embodying, e.g., also the purely hadronic case without a phase transition and to mimicking the occurrence of structures (so-called "pasta phases") in the phase transition region \cite{Yasutake:2014oxa,Alvarez-Castillo:2014dva}.

\section*{Acknowledgments}
%{\bf Acknowledgments}.
\noindent
This work was supported by NCN contract UMO-2014/13/B/ST9/02621
and by the COST Action MP1304 "NewCompStar". 
D.E.A-C., A.A. and H.G. received support from the Ter-Antonian-\-Smorodinsky and 
Bogoliubov-\-Infeld programmes. D.E.A-C. and S.T.  from the Heisenberg-Landau programme. 
A.A. acknowledges JINR grant No.~17-602-01.
S. B. acknowledges partial support by the Croatian Science Foundation under Project No. 8799.
D.B. was supported by the MEPhI Academic Excellence Project under contract No. 02.a03.21.0005 and
S.T. was supported by NAVI (VH-VI-417). 
%\bibliography{epja}

\end{document}